\documentclass{ws-procs9x6}

\newcommand{\refeq}[1]{(\ref{#1})}

\begin{document}

\title{SLOWLY ROTATING KERR BLACK HOLE AS A SOLUTION OF
EINSTEIN-CARTAN GRAVITY EXTENDED BY A CHERN-SIMONS TERM}

\author{\underline{M.\ CAMBIASO}}

\address{Instituto de Ciencias Nucleares,
Universidad Nacional Aut{\'o}noma de M{\'e}xico\\
A. Postal 70-543, 04510 M{\'e}xico D.F. and \\
Departamento de Ciencias F\'isicas, Universidad Andr\'es Bello \\
Rep\'ublica 220, Santiago, Chile\\
E-mail: mauro.cambiaso@nucleares.unam.mx}

\author{L.F.\ URRUTIA}

\address{Instituto de Ciencias Nucleares,
Universidad Nacional Aut{\'o}noma de M{\'e}xico\\
A. Postal 70-543, 04510 M{\'e}xico D.F.\\
E-mail: urrutia@nucleares.unam.mx}

\begin{abstract}
We consider the nondynamical Chern-Simons (CS) modification to
General Relativity (GR) in the framework  of the Einstein-Cartan
formulation, as providing a way to incorporate a  slowly
rotating Kerr black hole in the space of solutions.
Our proposal lies on considering the CS term as a
source of torsion and on an iterative procedure to look for
vacuum solutions of the system, by expanding the tetrad, the
connection and the embedding parameter, in powers of a
dimensionless small parameter $\beta$ which codifies  the
CS coupling. Starting from a torsionless zeroth-order
vacuum solution we derive the second-order differential equation
for the $\mathcal{O}(\beta)$  corrections to the metric, for an arbitrary
embedding parameter. Furthermore we can show that the slowly
rotating Kerr metric is an $\mathcal{O}(\beta)$ solution of the system
either in the canonical or the axial embeddings.
\end{abstract}

\bodymatter

\section{Introduction}
\label{sec:intro}
The CS modification of GR was introduced
in the framework of the Einstein-Hilbert
formulation of GR in Ref.\ \refcite{Jackiw:2003pm}.
The  model, referred
to as CS-EH, has been the subject of numerous investigations in the
literature\cite{Alexander:2009tp} and further understood
as providing a way to describe gravitational parity violation.
It can also be viewed as the GR analogue of the
Carroll-Field-Jackiw
version of extended electrodynamics by a Lorentz-symmetry breaking
term of the Chern-Simons kind,\cite{Carroll:1989vb} which
is one of the many terms of the mSME.
On the other hand it could be a means to implement a general
relativistic
version of Cohen and Glashow's Very Special
Relativity\cite{Cohen:2006ky}
after it was pointed out that the symmetries of
CS extended electrodynamics are those of one of the VSR
models.\cite{Hariton:2006zj}
Despite the many successes of this model and the fact that
some slowly rotating black hole solutions
have been
found,\cite{Konno:2007ze,Grumiller:2007rv, Konno:2009kg, Yunes:2009hc}
there remains an important unsolved question as to whether or not
the Kerr black hole is a solution of a CS modified gravity.

To this end we adhere to the viewpoint of treating  the CS extension of
GR in the Einstein-Cartan (EC) formulation as done by
Botta Cantcheff,\cite{Cantcheff:2008qn} which we will refer to as
CS-EC. Namely, the CS
coupling is implemented in terms of the SO(1,3)
gauge connection of GR, providing a source of torsion even in
the absence of matter,\cite{REVTORSION,Shapiro:2001rz} thus
putting torsion at the `forefront.'\cite{connection}

The action  reads:
\begin{equation}
S[e^{c},\omega^{a}{}_b]=\frac{\kappa }{2}\!\int \!\!\left( \mathbf{R}%
_{ab}\wedge ^{\ast }\!(\mathbf{e}^{a}\wedge \mathbf{e}^{b})+\beta \vartheta
\;\mathbf{R}^{a}{}_b \wedge \mathbf{R}^{b}{}_a \right) \!,
\label{ACTION}
\end{equation}%
where the Riemann $\mathbf{R}_{ab}$ and torsion $\mathbf{T}^{a}$ two-forms
are defined as:
\begin{equation}
\mathbf{R}^{a}{}_b=\mathbf{d\omega}^{a}{}_b+\mathbf{\omega }
^{a}{}_c \wedge \mathbf{\omega}^{c}{}_b\,,\;\;\;\quad \mathbf{T}%
^{a}=\mathbf{d}\mathbf{e}^{a}+\mathbf{\omega}^{a}{}_b\wedge \mathbf{e}%
^{b}\,.  \label{DEF R and T}
\end{equation}

In Eq.\ \refeq{ACTION} the variable  $\vartheta $, taken as
nondynamical, is the so called embedding parameter and $\beta $
is a  dimensionless parameter, which serves as a
bookkeeping device that will be clarified below.

The equations of motion for  Eq.\ \refeq{ACTION} are:
\begin{eqnarray}
\delta \mathbf{e}^{c} &:&\;\;\epsilon _{abcd}\mathbf{R}^{ab}(\omega )\wedge
\mathbf{e}^{c}=0\,,  \label{EINSTEIN} \\
\delta \mathbf{\omega }_{\;\;b}^{a} &:&\epsilon _{abqp}\;\mathbf{e}%
^{q}\wedge \mathbf{T}^{p}+2\beta \mathbf{d}\vartheta \;\wedge \mathbf{R}%
_{ab}(\omega )=0\,.  \label{TORSIONEQ}
\end{eqnarray}%

\section{Perturbation scheme of the CS term as a torsion source}
\label{sec:perturbation}
Equation \refeq{TORSIONEQ} reveals the role of the
CS term as a source of torsion and it can be solved
for the tensor components of the torsion tensor in terms of
the complete torsionful Riemann tensor to yield:
\begin{eqnarray}
T^{\sigma}{}_{\alpha \beta }&=&\frac{\beta }{2\, \mathrm{det}(e_\mu{}^a)}
(\partial _{\mu }\vartheta) %
\Big[ 2\epsilon ^{\mu \nu \rho \sigma }R_{\alpha \beta \nu \rho }
\nonumber\\
&&
\hskip 70pt
+ \left(
\delta _{\alpha }^{\sigma }\epsilon ^{\mu \nu \rho \omega }
R_{\beta \nu \rho
\omega }-\delta _{\beta }^{\sigma }
\epsilon ^{\mu \nu \rho \omega }R_{\alpha
\nu \rho \omega }\right) \Big].  \label{TEXACT}
\end{eqnarray}
Here $\epsilon ^{\mu \nu \rho \omega }=0, \pm 1$ is the
Levi-Civit\`a symbol. The LHS is in turn defined in terms
of the spin connection's components which also enter the
definition of the Riemann tensor; therefore the latter is a
highly non-linear equation for the spin connection, which we do not
aim to solve exactly for the moment.

Indeed, the severe observational constraints on the presence
of torsion in Nature,\cite{Kostelecky:2007kx} which in
our case is driven by the CS term, leads us to pose
the following ansatz for the dynamical variables:
\begin{equation}
e_{\mu}{}^{c}= e^{(0)}_\mu{}^c +\beta \, e^{(1)}_\mu{}^c
+\dots \,,\;\;\quad \omega _\mu{}^a{}_b
=\;\omega _{\;\;\;\mu \;\;\;b}^{(0)\;\;a}+\beta \,\omega
_{\;\;\;\mu \;\;\;b}^{(1)\;\;a}+\dots \,,
\label{EXP}
\end{equation}
and similarly for the embedding parameter.
All quantities such as the Riemann, torsion tensor, etc\dots
are expanded accordingly and zeroth order is
to be understood as torsion free.
Say we start from a
zeroth-order approximation which  might be taken as any vacuum
solution of Einstein equation
with zero torsion, thus providing $e^{(0)}, \omega^{(0)}, T^{(0)}=0$
and $R^{(0)}$. Then we obtain  $T^{(1)}$ according to
Eq.\ \refeq{TEXACT} and from $T^{(1)}$ we solve for
$\omega^{(1)}$, arising from Eq.\ \refeq{DEF R and T},
in terms of $\partial e^{(1)}$ plus zeroth-order
quantities including the zeroth-order covariant derivative.
Next we construct the first-order Riemann tensor,
according to Eq.\ \refeq{DEF R and T}.
This will introduce an additional derivative to
$\partial e^{(1)}$, such that the final equation
$R^{(1)}_{\mu\nu}=0$ will be of second order in the
unknown $e^{(1)}$, providing the first correction to
the metric $g^{(1)}_{\mu\nu}$, thus determining all first-order
variables. Second- and higher-order corrections
can be obtained iterating the procedure above.

\section{Consistence between dynamics and geometry}
\label{sec:consistence}
Field equations and  Bianchi identities together
result in consistency conditions that must be met by the
dynamical variables.
For lack of space these can only be briefly discussed here.  For example,
the field equations together impose
$\epsilon_a^{\phantom{a}bcd} T^a_{\phantom{a}bc} = 0$. Applying
the exterior covariant derivative $\mathcal{D}$ acting on
Lorentz tensor valued $p$-forms\cite{goeckeler} on the field
equations together with the Bianchi identities also written in terms of
$\mathcal{D}$ (e.g.,
$\mathcal{D} \wedge \mathbf{R}^a_{\phantom{a}b} = 0 $)
demands the resulting (torsionful) Ricci tensor
to be symmetric in its components, $R_{ab} = R_{ba}$. On the
other hand, the corresponding Einstein equation in tensorial
components reads $G^\mu_{\phantom{\mu}\nu} = 0$, where the
Einstein tensor is defined in terms of the torsionful Riemann
tensor and therefore $R^\mu_{\phantom{\mu}\nu} = 0$. Finally
from the first Bianchi identity commented above, we can form
$\epsilon _{abqp}\mathcal{D}\wedge (\mathbf{R}^{ab}(\omega
))\wedge \mathbf{e}^{c} = 0$, leading to:
\begin{equation}
\nabla _{\alpha } G^{\alpha}{}_{\psi }-T^{\theta
}{}_{\psi \beta }  R^{\beta}{}_\theta -\frac{1}{2}T^{\beta
}{}_{\alpha \theta }  R^{\alpha \theta}{}_{\beta \psi }=0\,,
\label{CONSIST1 AS KOSTELECKY}
\end{equation}%
which due to the vanishing of $R^\mu_{\phantom{\mu}\nu}$ implies:
\begin{equation}
T^{\beta}{}_{\alpha \theta } R^{\alpha \theta}{}_{\beta \psi }=0\,.
\label{CONSIST4}
\end{equation}%
We can only mention that, for the cases discussed
below, all the necessary
consistency conditions of the kind described above
are indeed satisfied
at least to the desired $\mathcal{O}(\beta)$ level.

\section{
Slowly rotating Kerr solution}
\label{sec:kerr}
Solving the field equations to first order in
$\beta$ as outlined above leads to:
\begin{eqnarray}
\frac{\vartheta^{(0)} _{;\lambda ;\alpha }}{e^{(0)}}\left( \epsilon
_{\;\;\;\;\;\mu }^{\lambda \delta \gamma }R_{\;\;\;\delta \gamma \nu
}^{(0)\;\;\;\;\alpha }+\epsilon _{\;\;\;\;\;\nu }^{\lambda \delta \gamma
}R_{\;\;\;\delta \gamma \mu }^{(0)\;\;\;\;\alpha }\right)
&=&
\nonumber\\
&&
\hskip -40pt
g_{\nu \alpha
\,;\mu }^{(1)\phantom{;\mu};\alpha }+g_{\mu \alpha \,;\nu }^{(1)%
\phantom{;\nu};\alpha }-g_{\mu \nu \,;\alpha }^{(1)\phantom{ ;\alpha};\alpha
}-g_{\phantom{(1)};\mu ;\nu }^{(1)}\, , \label{FINALEINSTEIN}
\end{eqnarray}%
where $g_{\mu \nu }^{(1)}\equiv e_{\mu \nu }^{(1)}+e_{\nu \mu }^{(1)}$,
$g^{(1)}\equiv g^{(0)\mu \nu }g_{\mu \nu }^{(1)}=2e_{\phantom{(1) \rho}\rho
}^{(1)\rho }$,  $e^{(0)} \equiv \mathrm{det} (e^{(0)}_\mu{}^a)$
and the semicolon denotes zeroth-order
covariant derivative.
It is verified that for the canonical embedding
$\vartheta^{(0)} = t/\mu$ and for $g^{(0)}_{\mu \nu}$ given by the
Schwarzschild geometry, the LHS of Eq.\ \refeq{FINALEINSTEIN}
vanishes and
$g^{(1)}_{\mu \nu} \equiv 0$ solves the remaining RHS.
Also, the consistency
conditions of Sec.\ \ref{sec:consistence} are satisfied
too at least to $\mathcal{O}(\beta)$.
Therefore
the Schwarzschild geometry is an $\mathcal{O}(\beta)$
solution of the theory.
However, though it is not a trivial solution, it can also
be verified that
$g_{\mu \nu}^{(1)} = -(2M^2/r) \sin^2\theta \, \delta^t_\mu \delta^\phi_\nu$
solves the equation and satisfies the consistency conditions too.
Thus we have found an $\mathcal{O}(\beta)$ consistent solution
which to first order reads
$g_{\mu \nu} = g^{(0)}_{\mu \nu} \,+\, \beta \,\, g^{(1)}_{\mu \nu}$.
Identifying $\beta \equiv a/M$, where $M$ is an spherical body's mass
and $a$ its angular momentum we can interpret the above solution
as describing a slowly rotating Kerr black hole if $a/M \ll 1$,
\begin{equation}
g_{\mu \nu} \approx g^{\mathrm{Schw}}_{\mu \nu}
- \frac{2Ma}{r}\sin^2\theta \, \delta^t_\mu \delta^\phi_\nu \, =\,
  g^{\mathrm{slow \,Kerr}}_{\mu \nu}\,.
\end{equation}

\section{Final comments}
\label{sec:comments}
In the language of the SME, the CS term $\vartheta$ can be viewed
as an externally prescribed explicit symmetry breaking quantity. Although this
may lead to inconsistencies pointed out in Ref.\ \refcite{Kostelecky:2003fs},
here our first concern is
to explore the consequences of taking torsion into account in CS extended
gravity. The case of a dynamical $\vartheta$ that undergoes spontaneous
symmetry breaking is clearly more interesting but will be dealt with elsewhere.
One of the important consequences that our investigation shows is the possibility
to accommodate the Kerr solution into the theory. This may be
considered as one of the flaws of the theory in its original
formulation, or at least a missing ingredient.
Thus our result can be considered as an important contribution in this field,
pointing towards the benefits of not limiting attention to spaces of solution
with a symmetric connection.

It is interesting to note that the above solution is still valid
for  $\vartheta^{(0)} = M r \cos \theta$, which we call an `axial'
embedding, producing a symmetry breaking direction parallel to the rotation
axis of the Kerr metric, i.e., suggesting that the
slowly rotating Kerr
black hole could be interpreted as arising from the breaking of spherical
to axial symmetry.

It is also true that the perturbative expansion here presented
may be valid only up to a given order, which needs to be understood
further. Nevertheless, this problem is not new in this context, for example, in a previous
formulation of CS-EC even the Schwarzschild solution was valid up to
a given order only.\cite{Cantcheff:2008qn}

\section*{Acknowledgments}
\label{sec:acknow}
Support from the grants DGAPA-UNAM-IN 111210 and CONACyT \#
55310 and useful comments
from R.\ Lehnert as well as discussions with  J.\ Alfaro and
H.\ Morales-T\'{e}cotl at the early stages of our investigation
in gravitational Chern-Simons theories are gratefully acknowledged.


\begin{thebibliography}{xx}

\bibitem{Jackiw:2003pm}
R.\ Jackiw and S.Y.\ Pi,
Phys.\ Rev.\ D \textbf{68}, 104012 (2003).

\bibitem{Alexander:2009tp}
S.\ Alexander and N.\ Yunes,
Phys.\ Rep.\ \textbf{480}, 1 (2009).

\bibitem{Carroll:1989vb}
  S.M.\ Carroll, G.B.\ Field and R.\ Jackiw,
  Phys.\ Rev.\  D {\bf 41}, 1231 (1990).

\bibitem{Cohen:2006ky}
  A.G.\ Cohen and S.L.\ Glashow,
  Phys.\ Rev.\ Lett.\  {\bf 97}, 021601 (2006).

\bibitem{Hariton:2006zj}
  A.J.\ Hariton and R.\ Lehnert,
  Phys.\ Lett.\  A {\bf 367}, 11 (2007).



\bibitem{Konno:2007ze}
K.\ Konno, T.\ Matsuyama and S.\ Tanda,
Phys.\ Rev.\ D \textbf{76}, 024009 (2007).

\bibitem{Grumiller:2007rv}
D.\ Grumiller and N.\ Yunes,
Phys.\ Rev.\ D \textbf{77}, 044015 (2008).

\bibitem{Konno:2009kg}
K.\ Konno, T.\ Matsuyama and S.\ Tanda,
Prog.\ Theor.\ Phys.\ \textbf{122}, 561 (2009).

\bibitem{Yunes:2009hc}
N.\ Yunes and F.\ Pretorius,
Phys.\ Rev.\ D \textbf{79}, 084043 (2009).

\bibitem{Cantcheff:2008qn}
M.\ Botta Cantcheff,
Phys.\ Rev.\ D \textbf{78}, 025002 (2008);
PoS {\bf IC2006}, 060 (2006).

\bibitem{REVTORSION}F.W.\ Hehl, J.D.\ McCrea, E.W.\ Mielke and Y.\ Ne'eman,
Phys.\ Rep.\ {\bf 258}, 1 (1995).

\bibitem{Shapiro:2001rz}
I.L.\ Shapiro,
  Phys.\ Rep.\  {\bf 357}, 113 (2002)


\bibitem{connection}
See J.\ Bjorken's and I.L.\ Shapiro's contributions to these Proceedings.


\bibitem{Kostelecky:2007kx}
V.A.\ Kosteleck\'y, N.\ Russell and J.\ Tasson,
Phys.\ Rev.\ Lett.\ \textbf{100}, 111102 (2008)

\bibitem{goeckeler}
M.\ Goeckeler and T.\ Schuecker,
{\it Differential Geometry, Gauge Theories, and Gravity,}
Cambridge University Press, 1987.

%
%
%
%

\bibitem{Kostelecky:2003fs}
V.A.\ Kosteleck\'y,
Phys.\ Rev.\ D \textbf{69}, 105009 (2004).





\end{thebibliography}
\end{document}